\documentclass[aps, pre, twocolumn, superscriptaddress, nofootinbib]{revtex4-2}

\usepackage{cmap}

\usepackage[english]{babel}
\usepackage{amsmath}
\usepackage{amssymb}
\usepackage{bm}
\usepackage[dvipsnames]{xcolor}

\usepackage{graphicx}
\usepackage[caption=false]{subfig}

\newcommand{\phantomsubfloat}[1]{
    {
        \captionsetup[subfigure]{labelformat=empty, labelfont=bf}
        \subfloat[][]{#1}
    }%
}

\usepackage[bookmarks=false]{hyperref}
\hypersetup{
    colorlinks=true,
    linkcolor=blue,
    citecolor=blue,
    urlcolor=blue
}

\begin{document}

\title{Metastable States in the $\bm {J_1 - J_2}$ Ising Model}

\author{ V. A. Abalmasov}
\email{abalmasov@iae.nsc.ru}
\affiliation{Institute of Automation and Electrometry SB RAS, 630090 Novosibirsk, Russia}

\author{ B. E. Vugmeister}
\email{vugmeister@datamir.net}
\affiliation{Datamir, Inc., Clifton, NJ 07012, USA}


\begin{abstract}

{We study the ${ J_1-J_2}$ Ising model on the square lattice using the random local field approximation (RLFA) and Monte Carlo (MC) simulations for various values of the ratio ${p = J_2/|J_1|}$ with antiferromagnetic coupling $J_2$, ensuring spin frustration. RLFA predicts metastable states with zero order parameter (polarization) at low temperature for ${p \in (0,1)}$. This is supported by our MC simulations, in which the system relaxes into metastable states with not only zero, but also with arbitrary polarization, depending on its initial value, external field, and temperature. We support our findings by calculating the energy barriers of these states at the level of individual spin flips relevant to the MC calculation. We discuss experimental conditions and compounds appropriate for experimental verification of our predictions.}

\end{abstract}

\maketitle

\renewcommand{\figurename}{{FIG.}}

\section{Introduction}

\begin{figure}[]
\centering
\includegraphics[width= 0.99 \columnwidth]{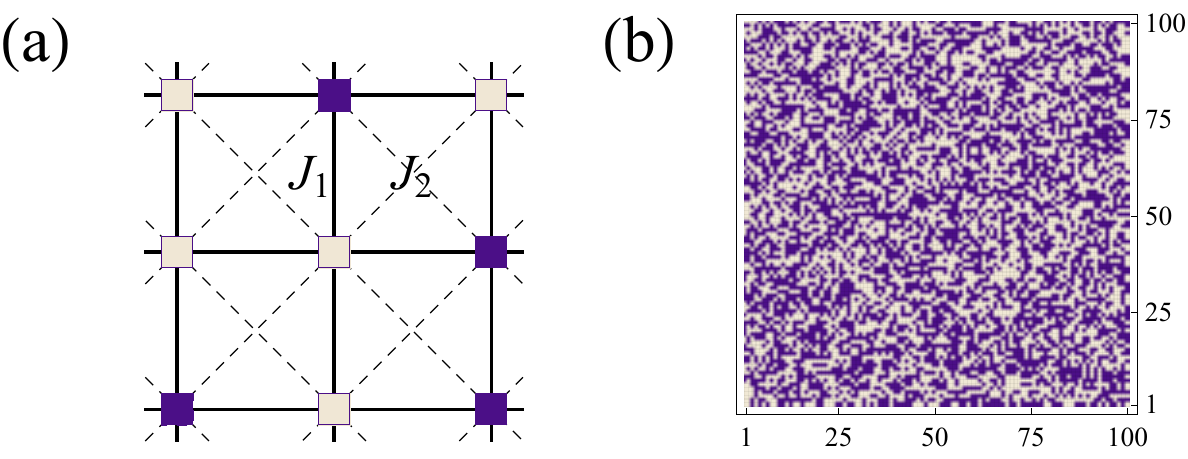}
\vspace{-0.cm}
\caption{{The ${J_1 - J_2}$ Ising model scheme.} {(a)},~Square lattice of Ising spins (white - up, blue - down) with the interaction constant $J_1$ between nearest neighbors along horizontal and vertical (solid) lines and $J_2$ between next-nearest-neighbors along (dashed) diagonals. {(b)},~Random (very high temperature) spin configuration; the number of spins along each side of the square sample is $L = 100$.}
\label{fig:1}
\end{figure}

In recent years, many compounds have been discovered in which electron spins form a two-dimensional square lattice and interact with their nearest and diagonal next-nearest neighbors via isotropic exchange interaction with the coupling constants $J_1$ and $J_2$, respectively (Fig.~\ref{fig:1})~\cite{ishikawa2017, mustonen2018a}. This also includes the parent compound of cuprate high-temperature superconductors La$_2$CuO$_4$~\cite{coldea2001, headings2010} and is likely relevant to iron-based superconductors \cite{si2016}. The corresponding quantum Heisenberg model has been studied extensively by a variety of methods, see e.g. \cite{manousakis1991, yu2012, wang2018} and references therein. For the values of the ratio $p = J_2/|J_1|$ near $p_0=1/2$, where two different ordered low  energy states have the same energy, the quantum spin liquid ground state was predicted~\cite{balents2010, jiang2012, ishikawa2017}, which may be the key to solving the problem of high-temperature superconductivity according to the resonating valence bond theory~\cite{anderson1987}. Indeed, this ground state has recently been observed experimentally in several compounds \cite{mustonen2018a, wen2019}.

The $J_1 - J_2$ Ising model, in which spins can only point in two directions, up and down, has also been thoroughly studied theoretically using the cluster mean field approximation~(MFA)~\cite{jin2013, bobak2015, kellermann2019}, Monte Carlo~(MC) simulations~\cite{kalz2011, jin2012, jin2013, ramazanov2016}, and tensor network simulations~\cite{hu2021, li2021}. Although its implementations seem less common in nature, the easier to study $J_1 - J_2$ Ising model nonetheless is interesting in its own right, and can also shed light on some properties of its more complex quantum Heisenberg counterpart. This is especially true for the Ising model in a transverse field, where quantum fluctuations induce gap between two ordered phases around $p_0$ \cite{sadrzadeh2016, bobak2018, kellermann2019, dominguez2021} with the valence-bond-solid state predicted \cite{sadrzadeh2016}. Indeed, the phase diagrams of both models have a lot in common~\cite{mustonen2018a, jiang2012, ishikawa2017, kellermann2019}.

Previously, it was shown that in the case of a ferromagnetic ground state, i.e. with the ferromagnetic~(FM) nearest-neighbor $J_1 = -1$ and antiferromagnetic~(AFM) next-nearest-neighbor $p \in (0, p_0)$ coupling constants, there exist local metastable states that slow down the dynamics of approaching equilibrium~\cite{shore1991, shore1992}. Later, specific metastable states were revealed in the 2D Ising model, i.e. at $p = 0$, during low temperature quenching with zero initial polarization in a zero external field~\cite{spirin2001a, spirin2001b, olejarz2011, olejarz2011b, olejarz2012}. In the present paper, we will extend the analysis of the features of metastable states for arbitrary values $p > 0$. First, we apply an analytical approach, the so-called random local field approximation (RLFA), to the problem. RLFA has been previously applied to reproduce the coexistence of ferrimagnetic order and cluster superparamagnetism in dilute spin systems, such as moderately impure lithium nickel superoxide compounds~\cite{mertz2001}. Its integral version~\cite{vugmeister1987} is particularly useful for long range spin-spin interactions and has been successfully applied to dilute Ising antiferromagnets~\cite{vugmeister1997} and relaxor ferroelectrics \cite{vugmeister1990}. We show that RLFA reveals metastable states with zero polarization at low temperature in zero field in the interval $p \in (0, 1)$. Note, however, that these metastable states are macroscopic states due to the nature of RLFA, which is an extension of MFA. At the same time, our MC simulations show that, in fact, metastable states can have arbitrary polarization values, and we discuss their nature and properties.

\section{Model and Methods} \label{model}

\subsection{Model}

Thus, we consider the Hamiltonian
\begin{align}\label{hamiltonianVug}
   H = J_1 \sum_{\langle i, j \rangle} s_i s_j + J_2 \sum_{\langle \langle i, j \rangle \rangle} s_i s_j - \sum_i h_i s_i,
\end{align}
where the sums are over nearest $\langle i, j \rangle$ and next-nearest (diagonal) $\langle \langle i, j \rangle \rangle$ neighbors (see Fig.~\ref{fig:1}), as well as over each spin coupled to the external field $h_i$ at its position with $s_i = \pm 1$. In what follows, we set FM $J_1 = -1$ and competing AFM $J_2 = p > 0$ coupling constants, which together ensure the frustration of the system. For this choice of couplings, the ground state is FM at $p < p_0$ and striped AFM at $p > p_0$; at $p = p_0$ the ground state is not ordered. The model properties remain the same for $J_1 > 0$ with the replacement of the uniform polarization by the N{\' e}el checkerboard one.

\subsection{Random Local Field Approximation}

The starting point of RLFA is the exact formula for the average spin value \cite{callen1963, vugmeister1987}:

\begin{align}\label{average-spin}
   \langle s_i \rangle = \langle \tanh \beta (h^s_i + h_i)\rangle,
\end{align}
where $\beta = 1/T$ is the inverse temperature (in energy units, with the Boltzmann constant set to unity, $k_B = 1$), $h^s_i = - \sum_j J_{ij} s_j$ is the local field acting on the spin $s_i$ due to all spins $s_j$ coupled with it; the brackets stand for the thermal averaging.

With RLFA the fluctuations of each spin are considered as independent, and averaging in Eq.~(\ref{average-spin}) is carried out with the product of the probability distributions for each spin \cite{vugmeister1987, mertz2001}:
\begin{align}   \label{probabilityM}
   P(s_i) &= (1 + m_i s_i)/2,
\end{align}
where $m_i = \langle s_i \rangle = m e^{i {\bf q} {\bf r}_i}$ is the thermally averaged polarization at the position ${\bf r}_i$ corresponding to the propagation vector~${\bf q}$. The uniform polarization corresponds to ${\bf q} = (0, 0)$, while the striped polarization refers to the vectors $(0, \pi)$ and $(\pi, 0)$. The same applies to the spatial dependence of the external field $h_i$.

Eqs.~(\ref{average-spin}) and (\ref{probabilityM}) together constitute the essence of RLFA.  Eq.~(\ref{average-spin}), which is a seventh degree polynomial in $m$ according to the eight neighbor spins in the model, can be solved numerically.

\subsection{Monte Carlo simulations}

We perform Monte Carlo simulations with single-spin-flip Glauber dynamics at zero temperature and Metropolis dynamics at low temperatures, making a deep quench from a (high-temperature) random or partially polarized initial spin configuration similar to what was done for the 2D Ising model in~\cite{spirin2001a, spirin2001b}, according to the following standard algorithm. If a randomly chosen spin flip leads to a negative energy change $\Delta E < 0$, then a new state is accepted. Otherwise, when $\Delta E \geq 0$, the probability of acceptance is $\alpha = \exp(-\Delta E /T)$ for the Metropolis algorithm, and $\alpha / (1 + \alpha)$ for the Glauber dynamics (zero temperature corresponds to the limit $T \rightarrow 0$). Both algorithms satisfy the detailed balance criteria and give the same final result~\cite{olejarz2011b, landau2009}.

\begin{figure*}[t]
\centering
\includegraphics[width=1. \textwidth]{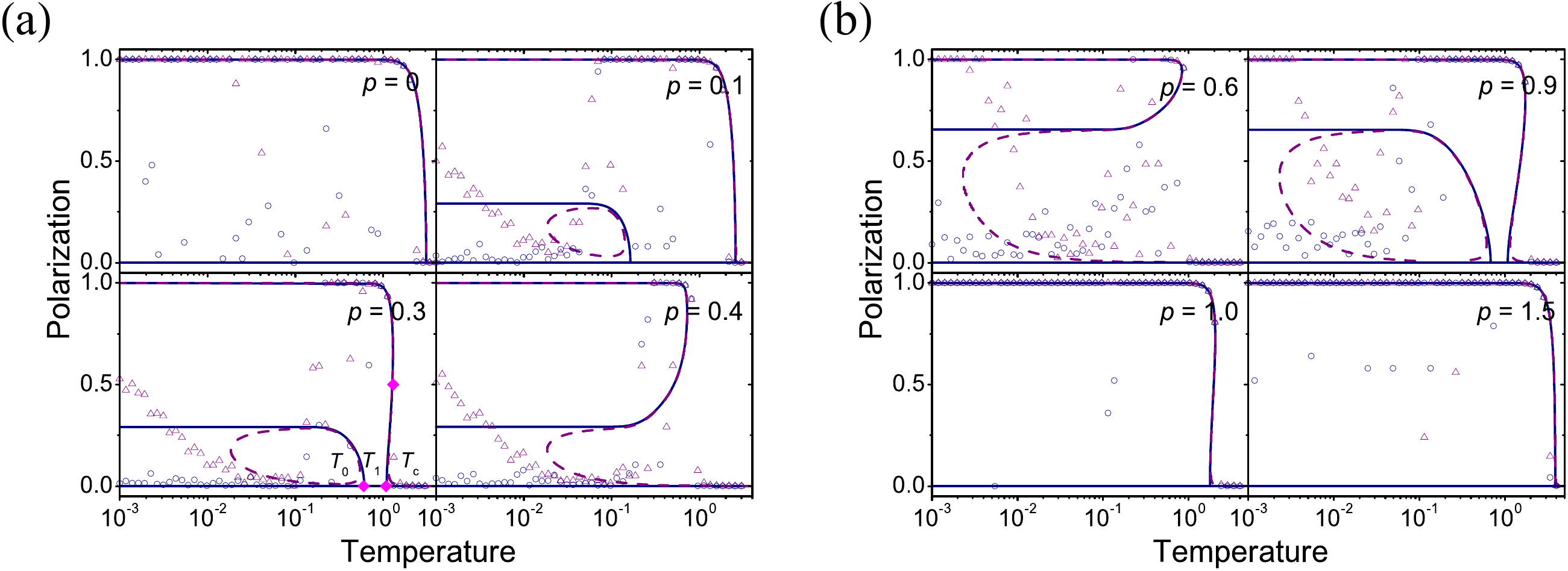}
\vspace{-1.2cm}
\flushleft
\phantomsubfloat{\label{fig:2a}}
\phantomsubfloat{\label{fig:2b}}
\caption{{Polarization as a function of temperature obtained using the RLFA analytical approach and MC simulations.} {(a)}, Solution of the RLFA equation (lines) and MC results (markers) for uniform polarization in a uniform field at $p < p_0$, and {(b)}, for striped polarization in a striped field at $p > p_0$. The magnitude of the field is $h = 0.001$ (purple dashed line and triangles) and $h = 0$ (blue solid line and circles). Each data point is derived from a single MC run with a random initial spin configuration at each temperature. The magenta diamonds in the bottom left panel (a) indicate the temperatures $T_0 < T_1 < T_c$. }
\label{fig:2}
\end{figure*}

We also obtain critical temperatures using MC to provide benchmarks for RLFA. They correspond to maxima of the susceptibility, which we calculate from the thermal fluctuations of the average value of $N$ spins $s = N ^{-1} \sum_{i=1}^N s_i e^{i {\bf q} {\bf r}_i}$, as $\chi = N T^{-1} (\langle s^2 \rangle - \langle s \rangle^2)$ with angular brackets representing thermal averaging~\cite{landau2009}.

\section{Results} \label{results}
\subsection{RLFA} \label{resultsRLFA}

The solution of the RLFA equation for uniform (at $p < p_0$) and striped (at $p > p_0$) polarization is shown in~Fig.~\ref{fig:2}. This solution corresponds to the zero value of the Landau potential derivative with respect to $m$ and, therefore, corresponds to its local minimum (stable or metastable state), local maximum or inflection point (unstable states). In the absence of an external field, zero polarization is always a solution to the equation, and it is unique and stable at high temperatures. At zero temperature, there is always (except in the case of $p = p_0$) another solution $m = 1$, which corresponds to full polarization and supposed to be stable. And there is still a third solution about $m_b \approx 0.29$ for $p \in (0, p_0)$ and $m_b \approx 0.65$ for $p \in (p_0, 1)$. It corresponds to a local maximum, i.e. barrier, of the Landau potential separating two local minima at $m = 1$ and $m = 0$, the first of which is a stable solution, and the second (with a higher energy) is a metastable one. In an external field, the metastable state exists only in a certain temperature window (dashed purple lines in Fig.~\ref{fig:2}) and completely disappears at fields above the critical $h_{\text{cr}}$. At $p$ = 0.1, 0.3, and 0.9, the values of this field are $h_\text{cr}$ = 0.0037, 0.0113, and 0.0373, respectively (the corresponding temperatures for the onset of metastable states are $T_0$ = 0.1634, 0.6083, and 0.6808). The barrier heights, which are the product $E_b = h_\text{cr} m_b$, are thus equal to 0.0011, 0.0033, and 0.0242, respectively.

\begin{figure*}[t]
\centering
\includegraphics[width=1. \textwidth]{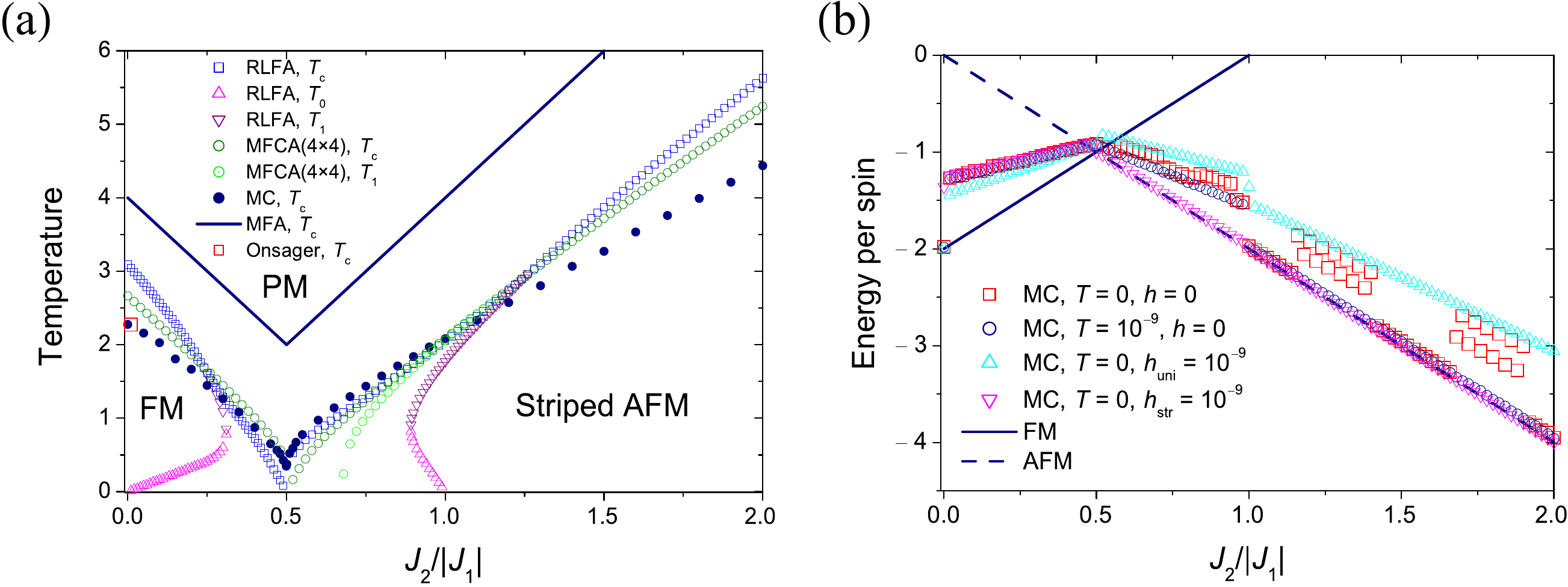}
\vspace{-1.4cm}
\flushleft
\phantomsubfloat{\label{fig:3a}}
\phantomsubfloat{\label{fig:3b}}
\caption{{Critical temperatures and energies in the ${J_1 - J_2}$ Ising model as functions of ${p = J_2/|J_1|}$.} {(a)}, Phase diagram obtained using RLFA, standard MFA and MC. Solid dark blue curves correspond to MFA, open blue circles, magenta up-triangles and purple down-triangles are $T_c$, $T_0$ and $T_1$ obtained within RLFA (Fig.~\ref{fig:2a}). The dependence $T_c(p)$ shows the logarithmic scaling in the striped phase as $p \rightarrow p_0$. The red square is the exact Onsager's solution for the 2D Ising model. Dark blue filled circles are calculated using the MC method. {(b)}, Energy per dipole obtained using MC simulation at zero temperature in the absence of a field (red squares), at a temperature $T = 10^{-9}$ and a field $h = 0$ (blue circles) and $T = 0$, $h = 10^{-9}$ (cyan up-triangles for uniform and magenta down-triangles for striped field). Each data point is averaged over 100 samples of size $L = 100$. Standard deviations of the energy distribution histogram over 100 samples for each data point are smaller than the markers. Blue solid and dashed lines correspond to the energy of the FM and AFM states, respectively. }
\label{fig:3}
\end{figure*}

The phase diagram for the $J_1 - J_2$ Ising model obtained using RLFA is shown in Fig.~\ref{fig:3a}, which also shows for comparison the critical temperatures from our MC simulation, which are in good agreement with the literature data~\cite{kalz2011, ramazanov2016}. In Fig.~\ref{fig:3a}, we also traced the temperatures $T_0$ and $T_1$ (see Fig.~\ref{fig:2a}, left bottom panel), which indicate the range of the zero-polarization metastable state and the lowest possible temperature of the first-order phase transition in the absence of an external field. The accuracy of RLFA regarding the values of $T_c$ turns out to be comparable to the cluster MFA~\cite{jin2013, bobak2015, kellermann2019}.

Within RLFA, the transition turns out to be first order for $p$ from about 0.25 up to 1.25, while it has recently been shown to be second order everywhere, except perhaps for the region $0.5 < g < 0.54$, using the tensor network simulation technique~\cite{li2021}. At the same time, the first order phase transition was also predicted just below $p_0$ with $T_c(p_0) = 0$ using the same method~\cite{hu2021}. It should be noted that the region of the first-order phase transition in the diagram narrowed as the quality of numerical calculations improved (see, for example, \cite{kalz2011, jin2012, jin2013, bobak2015, ramazanov2016}), and the behavior at the tricritical point $p_0$ was of particular interest~\cite{landau1981, rikvold1983, dasilva2013}. 

An important feature of the phase diagram, Fig.~\ref{fig:3a}, is that RLFA supports a logarithimic scaling of $T_c (p)$ as $p$ approaches $p_0$ from above. Indeed, in this region the RLFA equation (\ref{average-spin}) can be simplified under the assumptions $\Delta m = 1 - m \ll 1$ and $T \ll |J_1|, J_2$. Keeping the terms up to $\Delta m^2$ in the equation, from the unique solution condition (see Fig.~\ref{fig:2b}, $p = 0.6$) we find the equation $T_c = - 2 / [\ln(p - p_0) - \ln(1 + \sqrt{5}/2) - \ln T_c]$. The latter yields the dependence $T_c(p)$, which is very close to the asymptotic dependence $T_c (p) \sim -2.16 / \ln(p - p_0)$, obtained recently from the transfer matrix approach~\cite{hu2021}. However, as $p$ approaches $p_0$ from below, the RLFA solution tends to zero linearly, lowering $T_c$. 

It should be noted that neither striped for $p \in (0, p_0)$ nor uniform for $p \in (p_0, 1)$ polarizations with $m = 1$ are solutions to the RLFA equation, although they are very close to it.  At the same time,  these states are metastable at zero temperature, since any spin flip in these states leads to an increase in energy.

\subsection{MC numerical simulation}

\subsubsection{Metastable states}

To further explore the metastable states, we perform MC simulations with single-spin-flip dynamics, making a deep quench from a (high-temperature) random spin configuration. Note that using the same technique, it was previously shown~\cite{spirin2001a, spirin2001b} that at zero temperature, starting from zero polarization, approximately in every third case the 2D Ising system (p=0) reaches a metastable state with a pair of vertical or horizontal stripes. This was later explained by revealing a deep connection between the zero-temperature coarsening with critical continuum percolation \cite{olejarz2012, blanchard2013}. However, for a finite polarization or any external field, the 2D Ising system always reaches ground states at zero temperature. We will show below that quite different situation occurs in the $J_1-J_2$ Ising model for $p>0$.

In all calculations, we use the sample size $L = 100$ (unless otherwise specified) and periodic boundary conditions. To obtain the data in Fig.~\ref{fig:2}, relaxation is performed at each temperature, starting from a (high-temperature) random spin configuration, with $10^5$ Monte Carlo steps per spin (MCS) used for thermalization and the same number of MCS for subsequent calculations of thermodynamic quantities for each run. This is much more than the domain growth time of about $L^2$ in units of MCS required to reach the ground state after quenching in the absence of metastable states \cite{sadiq1984, cirillo1997} (note that the diagonal domain growth observed in the Ising model is longer and requires about $L^3$, but this only occurs about 4\% of the time \cite{spirin2001a}). Metastable states, in turn, are reached in a much shorter time of about 100 MCS (see below), and their further relaxation is determined by the energy barrier together with the thermal activation law and is very long at low temperature. When simulating at zero temperature (Fig.~\ref{fig:3b}), each data point is averaged over 100 samples with a relaxation time of~$10^4$~MCS. For striped polarization, the largest of the values corresponding to the propagation vectors ${\bf q} = (0, \pi)$ and $(\pi, 0)$ is taken.

\begin{figure*}[t]
\centering
\includegraphics[width=1. \textwidth]{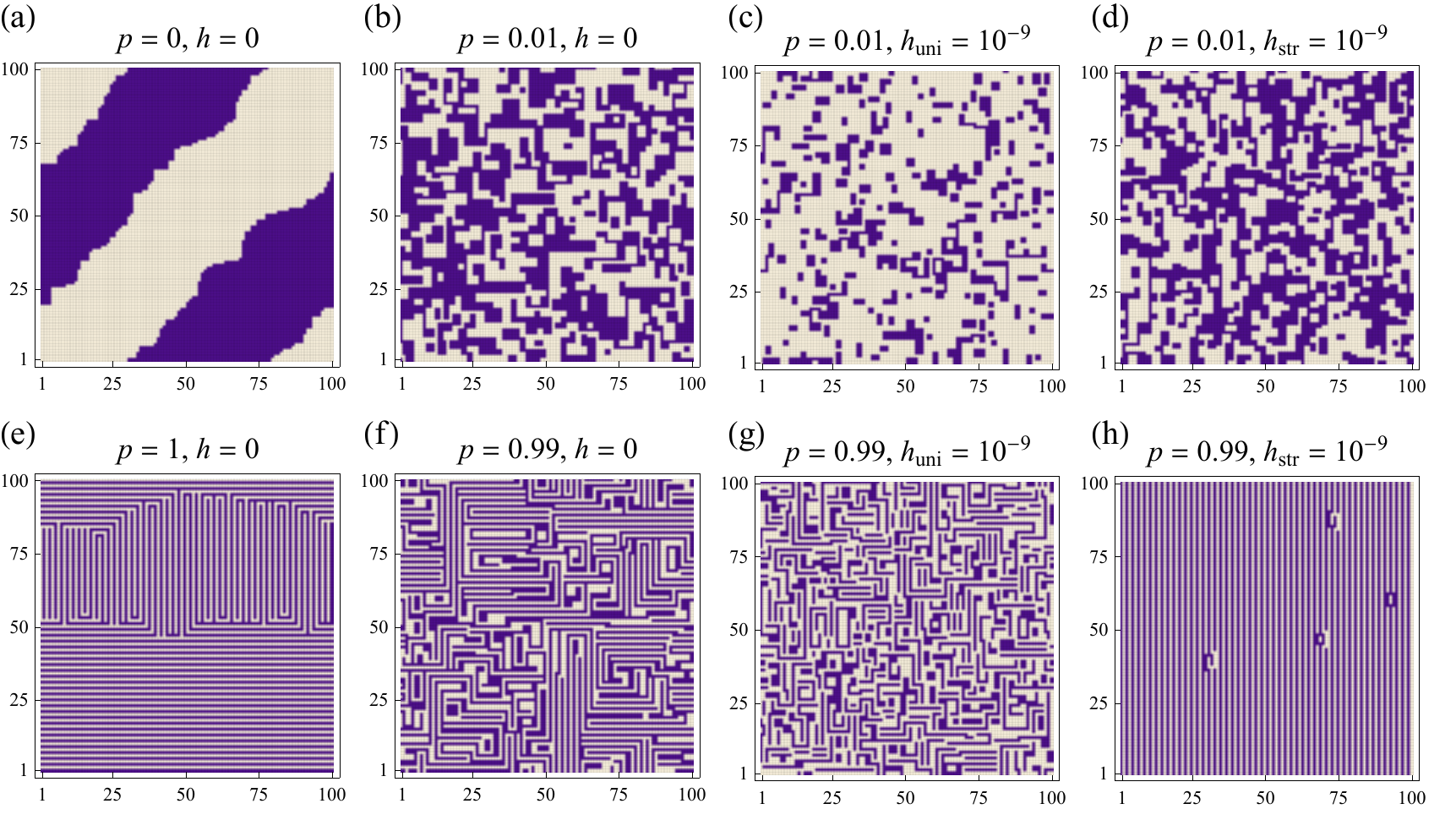}
\vspace{-1.4cm}
\flushleft
\phantomsubfloat{\label{fig:4a}}
\phantomsubfloat{\label{fig:4b}}
\phantomsubfloat{\label{fig:4c}}
\phantomsubfloat{\label{fig:4d}}
\phantomsubfloat{\label{fig:4e}}
\phantomsubfloat{\label{fig:4f}}
\phantomsubfloat{\label{fig:4g}}
\phantomsubfloat{\label{fig:4h}}
\caption{Spin configurations of final absorbing states after energy relaxation at zero temperature, starting from a random spin configuration, in the absence or in a very small uniform or striped external field $h$ for the values of $p = J_2/|J_1|$ equal to 0, 0.01, 0.99, and 1.}
\label{fig:4}
\end{figure*}

In the absence of an external field, the MC simulations show critical slowdown, which corresponds to the metastable states with nearly zero polarization for $p \in (0, 1)$. The temperature below which the initial state remains frozen is linear in the activation energy of the metastable state and depends logarthmically on the MC relaxation time. It is well below what follows from RLFA~(Fig.~\ref{fig:2}). Typical spin configurations of these states at zero temperature are shown in Figs.~\ref{fig:4b}, ~\ref{fig:4f}. The result is the same when the initial spin configuration at each temperature, instead of a random one, is AFM for $p \in (0, p_0)$ or FM for $p \in (p_0, 1)$.

For a small external field $h$ (uniform for $p < p_0$ and striped for $p > p_0$) at temperatures $T < h$,  relaxation gets stuck in metastable states with higher polarization~(Fig.~\ref{fig:2}). At zero temperature, these metastable states appear already in an infinitesimally low external field (e.g. $10^{-9}$ as in our simulations) with their typical configurations shown in Figs.~\ref{fig:4c}, ~\ref{fig:4h}. The polarization of these states is about $m \approx 0.56$ for $p < p_0$, and $m \approx 0.98$ for $p > p_0$. As the temperature increases for $p < p_0$, the polarization first decreases and then increases to $m = 1$ before the FM phase transition. For $p > p_0$, however, at $T \approx h$ the polarization $m  = 1$ is observed after quenching, and only at $T > h$ does it decrease and increase similarly to the case $p < p_0$. When we swap the external field propagation vectors, the metastable states configurations at $T = 0$ become somewhat denser compared to the zero field case, see~Fig.~\ref{fig:4d} and~Fig.~\ref{fig:4g}.

We note that some data with an intermediate polarization value in Fig.~\ref{fig:2} actually correspond to incompletely relaxed FM and AFM states divided into slowly relaxing large diagonal domains like in Figs.~\ref{fig:4a} or nondiagonal metastable stripes (see the discussion about these states for the case of $p = 0$ in~\cite{spirin2001a, spirin2001b}). At the same time, the energy of these states does not differ much from a completely ordered state, while for truly disordered states it is noticeably higher. Thus, we plot the energy per spin after relaxation of random initial spin configurations at zero temperature as a function of $p$ (Fig.~\ref{fig:3b}), where the metastable states at $p \in (0,1)$ (red squares) are clearly visible. For some values of $p > p_0$, the energy of metastable states in Fig.~\ref{fig:3b} appears to be slightly higher and goes above the general trend. However, it suffices to apply an infinitesimal temperature, for example, $T \sim 10^{-9}$, as in our simulation, for these fragile metastable states to quickly relax into robust metastable states for $p \in (p_0, 1)$ or stable states for $p > 1$. Thus, these states, which have a mosaic domain structure and appear at each run, resemble metastable states with horizontal and vertical stripes in the 2D Ising model, appearing at about every third quenching~\cite{spirin2001a, spirin2001b}. The energy deviations of these metastable states is less than 0.05, and changing the sample size to $L = 50$ does not change the diagram in~Fig.~\ref{fig:3b}.

\subsubsection{Approaching equilibrium}
\label{resultsMCDynamic}

In contrast to the 2D Ising model (p = 0)~\cite{spirin2001a, spirin2001b}, for $p \in (0,1)$ and $T = 0$ the system always gets stuck in a metastable state upon quenching. But similarly, the relaxation time of metastable states to the ground state is determined by the activation energy $E_a$ as $\tau \propto \exp(E_a/T)$~\cite{spirin2001a, spirin2001b}. For $p \in (0, p_0)$, where the ground state is FM, metastable states consist of rectangles with at least two spins on each side, surrounded by spins with the opposite direction. These rectangles are then interconnected, making up the whole picture (Figs.~\ref{fig:4b} - \ref{fig:4d}). The energy cost for a spin flip in the corner of the rectangle is $4 J_2$, on its side is $4 J_1$, and in the middle of a long line of one spin wide is $8 J_2$. The minimum of these energies yields the activation energy $E_a = 4 J_2$ (see \cite{shore1991, shore1992} for a similar discussion). The external field sufficient to flip a spin and destroy the metastable state is half of this value and is equal to $h_a = 2 J_2$. For $p \in (p_0, 1)$, the simplest excitations appear to be rectangles three spins wide, more than four spins long, and of opposite polarization inside~(Fig.~\ref{fig:4h}). The activation energy must be a minimum of $4 (J_1+2 J_2)$ for spins inside and $-4 (J_1+J_2)$ at the border of the rectangle.

\begin{figure*}[]
\centering
\includegraphics[width=1. \textwidth]{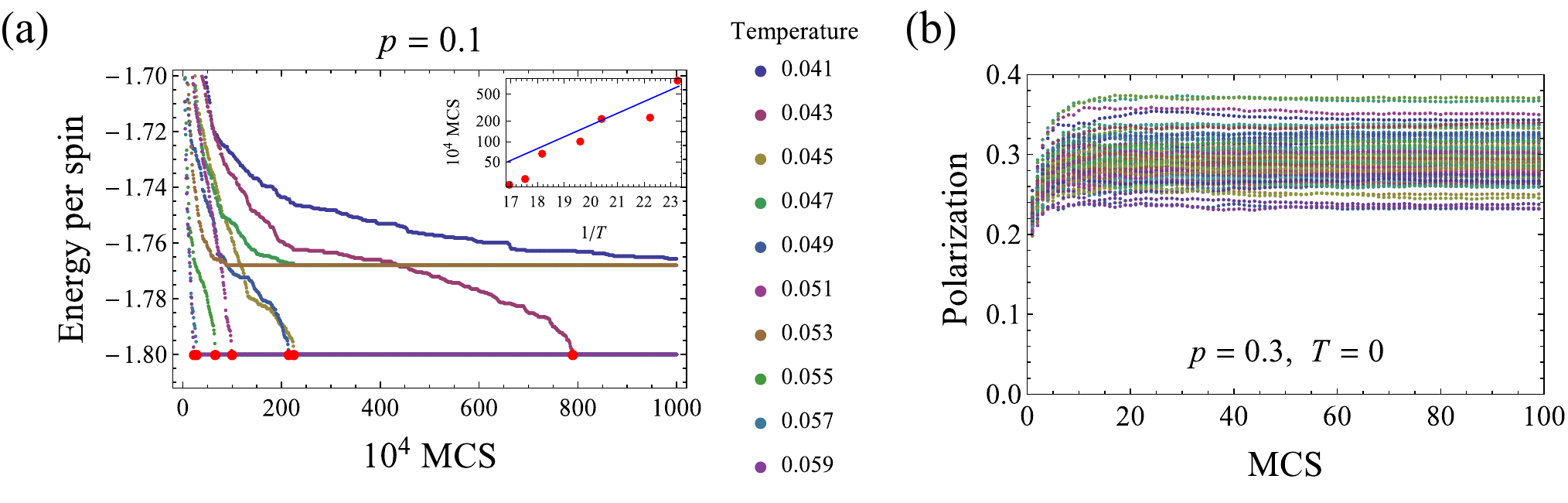}
\phantomsubfloat{\label{fig:5a}}
\phantomsubfloat{\label{fig:5b}}
\caption{Relaxation during MC simulation. (a), Energy relaxation starting from a random spin configuration at each temperature at $p = 0.1$, $L =100$. In some cases, the system is stuck in a state with an energy of about $E_1 = -1.77$, which is higher than the FM ground state with $E_0 = -1.8$ by the energy of two domain walls. These two-stripe states are metastable with an activation energy of $4|J_1|$ and disappear in an external field, as discussed for the 2D Ising model in~\cite{spirin2001a, spirin2001b}. Large red dots at energy $E_0$ correspond to the relaxation time. Inset: the relaxation time fitting to the activation law $A \exp(0.4/T)$ vs $1/T$.  (b), Polarization relaxation at $p = 0.3$ and zero temperature for 100 samples, starting from a random spin configuration with polarization $m_{\text{init}} = 0.2$.}
\label{fig:5}
\end{figure*}

Our MC calculation in an external field at zero temperature confirms the critical fields $h_a$ given above. At temperatures below $T_c$, where the critical slowing down due to the phase transition is negligible, relaxation to the ground state is fast enough with a relaxation time of the order of hundreds of MCS. At still lower temperatures, after a very quick relaxation to the metastable state over tens of MCS, the system begins to slowly relax to the ground state. The corresponding relaxation time can be derived in a narrow temperature window, see~Fig.~\ref{fig:5a}, and it appears to be in accordance with the activation law supporting the activation energies written above. For example, at $p = 0.1$ and $L = 100$, a least square fit with $A \exp(0.4/T)$ yields $A = 590$, see~Fig.~\ref{fig:5a}. The same activation law was obtained for small $L \leq 8$ in~\cite{shore1992}.

When we start relaxation from a random spin configuration with nonzero polarization at low temperatures and $p \in (0, 1)$, the resulting state is not the ground state, as in the case of $p = 0$~\cite{spirin2001a, spirin2001b}, but also has a nonzero polarization, slightly higher than the initial one, see Fig.~\ref{fig:5b}. This proves the existence of metastable states with an arbitrary polarization value in the $J_1-J_2$ Ising model.

The MC calculations also show that for nonzero $p$, even at zero temperature, metastable states can fluctuate between different configurations, slightly changing their polarization but keeping their energy constant. This effect was previously observed in the Ising model in a 3D cubic lattice~\cite{spirin2001a, spirin2001b, olejarz2011, olejarz2011b} and some 2D lattices~\cite{yu2017}, and these states were called blinking states. This occurs when the local field $h_i^s$ on the spin, due to its nearest and next-nearest neighbors, is zero. Any external field, however, removes this degeneracy, and blinking states disappear. For $p = 1$, a blinking state is shown in Fig.~\ref{fig:4e}.

\section{Discussion}

The main result of this work is the discovery of specific features of metastable states at $p \in (0, 1)$, analyzed by RLFA and MC simulations. It means that the spin frustration, which is present for all $p > 0$, is not a sufficient condition for the appearance of metastable states in this case. Only fragile metastable states have been revealed at certain values of $p > 1$ using MC.

The results of RLFA should not be understood literally. Being just a modification of MFA, it can predict only some general trends. Thus, it predicts a metastable state with zero polarization (in zero external field), while the MC simulation reveals the presence of metastable states with arbitrary polarization values. Within RLFA, metastable states appear only below a certain temperature $T_0$, while from a microscopic point of view these states exist at all temperatures, but the system effectively get stuck in them when the temperature is much less than the activation energy $E_a$ to exit them, see~Fig.~\ref{fig:5a}. 

Although the temperatures $T_0$, below which metastable states appear in RLFA, are of the order of the activation energies $E_a$ from MC, the critical fields $h_a$ found from the MC calculation turn out to be much larger than the values obtained within RLFA. Given the mean-field nature of RLFA, one can assume that its critical field is the average value for all spins in the sample. Indeed, with a total number of spin configurations around the central spin of $2^8$, only $4$ of them have the central spin in a rectangular corner that is part of a metastable domain. Thus, we can estimate that the critical fields in RLFA are 64 times smaller than those calculated microscopically for a spin flip trapped in a metastable state. This is consistent with the value of about 54 of the ratio of critical fields calculated from the formula obtained in~Sec.~\ref{resultsMCDynamic} to the RLFA values of $h_\text{cr}$ given in~Sec.~\ref{resultsRLFA} for $p= 0.1$ and 0.3.

External fields $h$ significantly affect spin quenching in the MC simulation. At zero temperature and fields above the critical one, $h > h_a$, the system relaxes to the ground state. When $h < h_a$ and $T < h$, the system relaxes to a state with polarization $m < 1$~(Fig.~\ref{fig:2}), while RLFA predicts relaxation to the ground state. This effect should be due to the destruction of fragile metastable states in an external field, similar to what was observed in the 2D Ising model~\cite{spirin2001a, spirin2001b}. On the other hand, at temperatures $h < T < h_a$, the system relaxes to a state with a polarization close to the initial one, just as in zero field. In this case, the external field is smaller than any energy scale and effectively vanishes in the problem, and the RLFA and MC results are in qualitative agreement~(Fig.~\ref{fig:2}).

As an alternative, we also used the mean-field cluster approximation with the $4\!\times\!4$ cluster size, formulated as in \cite{blinc1966b, abalmassov2019}, but, as it turned out, it does not predict metastable states. Although a $4\!\times\!4$ cluster size is sufficient to study the $J_1-J_2$ Ising model in many cases~\cite{jin2013, bobak2015, kellermann2019}, it is possible that a larger cluster is needed to reveal metastable states. For instance, metastable states in an external field near critical points have recently been predicted for the Ising model with competing long-range interactions using the $6\!\times\!6$ cluster approximation~\cite{rikvold2016, nishino2018}.

According to the calculated phase diagram (Fig.~\ref{fig:3a}), the accuracy of RLFA turned out to be similar to the accuracy of the cluster MFA ordinary applied to this problem \cite{jin2013, bobak2015, kellermann2019}. Thus, besides metastable states, RLFA is also an effective tool for studying phase diagrams in general, and it would be promising to use it, for example, to reproduce recently discovered anomalies in the dipole ordering of water molecules in minerals~\cite{belyanchikov2020, abalmasov2021a, belyanchikov2022, chan2022}, or to apply it to the above-mentioned Ising model with competing long-range interactions~\cite{rikvold2016, nishino2018}.

A legitimate question arises: To what extent one can expand the finding of the existence of metastable states in the $J_1-J_2$ Ising model to a broad and practically important class of compounds with $J_1-J_2$ Heisenberg spin interactions related to high temperature superconductors? Apparently, a first step in answering this question would be to investigate the Ising model in a transverse field. The latter has some common features with the quantum Heisenberg model, since both have non-diagonal fluctuating terms in the Hamiltonian, which may be factor in destroying metastable states. Our preliminary application of RLFA to the Ising model favors the existence of metastable states in sufficiently small transverse fields. On the other hand, by listing the appropriate compounds with $J_1-J_2$ Heisenberg interaction we encourage experimenters to look for possible metastable states in the experiments.

Turning to materials in which metastable states can exist, we first mention La$_2$CuO$_4$ with the N{\' e}el AFM ground state, in which $J_1 \approx 150$~meV was obtained from inelastic neutron scattering data  \cite{coldea2001, headings2010} in general  agreement with ab-initio calculations \cite{annett1989, katanin2002, wan2009, yamamoto2019}. However, the value and the sign of $J_2$ differ in different sources with $J_2 < 0$ in \cite{coldea2001, wan2009} and $J_2$ ranging from about 0.2 \cite{katanin2002, headings2010, yamamoto2019} to 0.8 in \cite{annett1989}.

For iron-based superconductors, which have a striped AFM ground state in their parent compounds \cite{dai2015, bascones2016}, it was shown that biquadratic coupling together with isotropic in-plane coupling constants explain many of the observed features \cite{wysocki2011, glasbrenner2014}. For CaFe$_2$As$_2$, for example, experimental data are well fitted for the ratio $p = 0.86$~\cite{wysocki2011}. We also mention LaFeAsO, where $p$ is very slightly more than one as calculated in \cite{ma2008}, while it was claimed about 0.71 in \cite{yildirim2008, yaresko2009}. In both compounds $J_1 > 0$ and they could be tested as well as La$_2$CuO$_4$ for metastable states.

Other suitable magnetic compounds corresponding to the $J_1-J_2$ Heisenberg model also include VOMoO$_4$ with $J_1 = 100 - 150$~K and $p \simeq 0.2$, and the N{\' e}el temperature $T_N = 42$~K \cite{carretta2002, bombardi2005}. In BaCdVO(PO$_4$)$_2$, the ground state is striped AFM with $J_1 = -3.6$~K and $J_2 = 3.2$~K \cite{nath2008, povarov2019}, which gives $p \simeq 0.9$. However, the expected temperature of metastable states is approximately two orders of magnitude lower (Fig.~\ref{fig:2}) than the already low phase transition temperature $T_N = 1.05$~K~\cite{povarov2019}, which may complicate its experimental study. In PbVO$_3$, where $J_1 \approx 190 - 200$~K and $p \approx 0.2 - 0.4$ is close to the gap around $p_0$ in the phase diagram, there is no long-range magnetic ordering down to 1.8 K~\cite{tsirlin2008}. The solid solution  Sr$_2$Cu(Te$_{1-x}$W$_x$)O$_6$ is unique for studying frustrated square-lattice antiferromagnetism as it can be tuned from the N{\' e}el ($x = 0$, $J_1 \approx 83$~K, $p \approx 0.03$) to the striped AFM order ($x = 1$, $J_1 \approx 14$~K, $p \approx 7.92$) by varying the composition~\cite{mustonen2018b}. Thus, this compound may also be a preferred choice for studying metastable states.

Experimentally, strongly nonequilibrium conditions equivalent to quenching can be achieved in ultrafast light-pump experiments, such as~\cite{afanasiev2021}. Note that metastable states have recently been reported in incipient ferroelectric SrTiO$_3$ under high-intensity THz pumping~\cite{li2019, nova2019} and were predicted from {\it ab-initio} calculations in the antiferroelectric NaNbO$_3$~\cite{prosandeev2022}. The nonequilibrium conditions can also be created by applying an external field at low temperature and turning it off abruptly. The switch-off time in this case must be less than the spin relaxation time, while in electrical circuits it is limited to hundreds of microseconds for magnetic fields~\cite{dedman2001}, and hundreds of picoseconds for electric fields~\cite{li2004}. However, in experiments with laser pumping, it can be short enough in both cases~\cite{vahaplar2009, mankowsky2017, abalmasov2020, abalmasov2021b}.

\section{Conclusion}

In this work, using RLFA, we predict the existence of metastable states with zero polarization in the $J_1-J_2$ Ising model at low temperature for $p \in (0, 1)$. Our MC simulations also indicate metastable states with an arbitrary polarization value. The energy barrier of these states depends on the coupling constant $J_2$. We point to some antiferromagnets, including known high-temperature superconductors, where these states could be expected at low temperature. These findings may be crucial for explaining the magnetic and electric properties of some materials and may directly manifest themselves, in particular, under the nonequilibrium conditions of modern experiments with high-power ultrashort light pumping.

\section*{Acknowledgments}

V.A.A. acknowledges the support by the Ministry of Science and Higher Education of the Russian Federation (No. 121032400052-6). The Siberian Branch of the Russian Academy of Sciences (SB RAS) Siberian Supercomputer Center is gratefully acknowledged for providing supercomputer facilities.



%

\end{document}